\documentclass[prl,twocolumn,longbibliography,superscriptaddress]{revtex4-1}
\usepackage{amssymb,amsmath}
\usepackage{xcolor}

\DeclareMathOperator{\Tr}{Tr}

\DeclareMathOperator{\sign}{sign}
\usepackage{float}
\usepackage{upgreek}

\usepackage[english]{babel}
\usepackage{graphicx}
\usepackage{verbatim}
\usepackage[colorlinks,urlcolor=blue,citecolor=blue,linkcolor=blue]{hyperref}
\usepackage{cancel}

\renewcommand{\vec}[1]{\mathbf{#1}}
\newcommand{\Frac}[2]{\displaystyle\frac{#1}{#2}}

\renewcommand{\k}{{\bf k}}
\newcommand{\x}{{\bf x}}
\newcommand{\p}{{\bf p}}

\newcommand{\q}{{\bf q}}
\newcommand{\Q}{{\bf Q}}
\newcommand{\0}{{\bf 0}}
\newcommand{\A}{{\bf A}}

\newcommand{\units}[1]{{#1}}

\newcommand{\aB}{a_{\text{B}}}

\newcommand{\EB}{E_{\text{B}}}

\graphicspath{{Figures/}}

\begin{document}
%\title{Charge-imbalanced polariton condensates}
\title{Crescent states in charge-imbalanced polariton condensates} 
\author{Artem Strashko}
\affiliation{Center for Computational Quantum Physics, Flatiron Institute, 162 5th Avenue, New York, NY 10010, USA}
\author{Francesca  M. Marchetti}
\affiliation{Departamento de Fisica Teorica de la Materia Condensada \& Condensed Matter Physics Center (IFIMAC), Universidad Autonoma de Madrid, Madrid 28049, Spain}
\author{Allan H. MacDonald}
\affiliation{Department of Physics, University of Texas, Austin, Texas 78712, USA}
\author{Jonathan Keeling}
\affiliation{SUPA, School of Physics and Astronomy, University of St Andrews, St Andrews, KY16 9SS, United Kingdom}
\date{\today}

\begin{abstract}
 We study two-dimensional charge-imbalanced electron-hole systems embedded in an optical microcavity. 
 We find that strong coupling to  photons favors  states with pairing at zero or small center 
 of mass momentum, leading to a condensed state with spontaneously broken time-reversal and 
 rotational symmetry, and unpaired carriers that occupy
 an anisotropic crescent-shaped sliver of momentum space. 
 The crescent state is favoured at moderate charge imbalance, while a Fulde--Ferrel--Larkin--Ovchinnikov-like state --- with  pairing at large center of mass momentum  --- occurs instead at strong imbalance. 
 The crescent state stability results from long-range Coulomb interactions in combination with 
 extremely long-range photon-mediated
 interactions. 
\end{abstract}

\maketitle

\emph{Introduction---}
At low-carrier densities, electrons and holes in two-dimensional semiconductors pair into bosonic excitons that can condense at low enough temperatures~\cite{keldysh1964possible,comte82:exciton1,high2012spontaneous,fogler2014high,Wang_Nature_2019}. Exciton condensation is expected to survive the frustration of unequal electron and hole densities~\cite{Pieri2007,Subashi2010,Kazuo2010,Parish_EPL2011,Varley_Lee_PRB2016}, which favors 
condensed electron-hole pairs that acquire a finite centre-of-mass momentum forming a state similar to the Fulde--Ferrel~\cite{Fulde-Ferrell_PR1964} (FF) and Larkin--Ovchinnikov~\cite{Larkin-Ovchinnikov_1964} (LO) phases (abbreviated as FFLO) known from superconductors.  
The prospect of FFLO phases has also been extensively discussed in the context of cold atoms~\cite{sheehy2007bec}. 
Although FFLO phases are common to imbalanced two-component fermions  with attractive interactions, more exotic alternatives, such as phase separation in momentum space (also named ``breached pair'' or ``Sarma'' phases) have been suggested in special cases~\cite{sarma1963influence,Forbes2005:BP}. 
In neutral systems, these uniform density 
imbalanced phases compete with, and are largely replaced by, phase separation in real space~\cite{zwierlein2006fermionic,partridge2006pairing,shin2006observation}.
For the charged electron-hole systems we focus on here, however, the electrostatic energy forbids phase separation and exotic uniform states are a stronger possibility. 

\begin{figure}[!htpb]  
\includegraphics[width=1.0\linewidth]{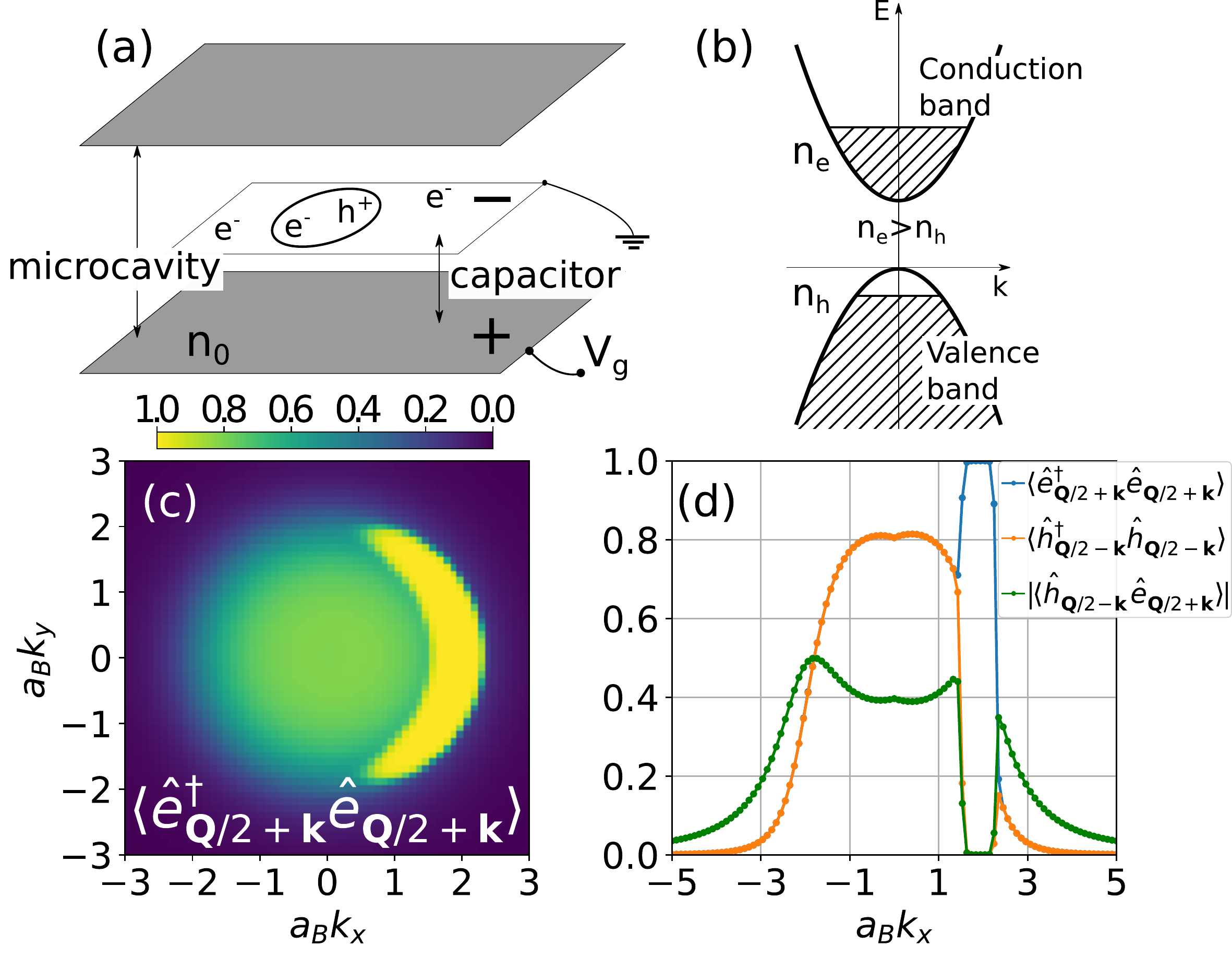}
\caption{(a) Semiconductor quantum well embedded in a planar microcavity, with net charge tuned by a gate 
voltage between the bottom  mirror and the grounded semiconductor.  (b) Occupied bands with finite excitation and 
charge. (c) Typical anisotropic crescent state, represented by the electron occupation
numbers, which
reaches one at low temperatures inside the yellow crescent-shaped region.
(d) $k_y=0$ momentum space slice of (c), showing both occupations and electron-hole 
coherence.  Inside the Fermi surface (yellow in (c)),
both conduction and valence bands are 
occupied so coherence vanishes.  Elsewhere in momentum space only one state is 
occupied.   
Results were calculated using the model parameters explained in the   text: target charge density $n_0 = 8.125 \times 10^{-2} \aB^{-2}$, excitation chemical potential relative to band gap
$\mu_{ex}-E_G = \EB$, temperature $k_B T = 0.04\EB$, photon cutoff frequency $\omega_0 = 3.06\EB$, matter-light coupling momentum cutoff $\kappa = 2.5  \aB^{-1}$, matter-light coupling $g_0 = 0.8 \EB \aB$,  mass ratio $m_e/m_h=1$, $\varepsilon = 1$, and  capacitive energy $\alpha = 800 \EB \aB^2$.
}
\label{fig:cartoon}
\end{figure}

The boson condensation temperature increases significantly when optically-pumped two-dimensional 
semiconductors are placed in a planar  microcavity,  designed so that long-wavelength confined photons are  close to resonance with excitons~\cite{Kasprzak_Nature2006,Balili2007a}.  The resulting quasiparticles, exciton-polaritons, are photon--exciton hybrids, that have a greatly reduced mass~\cite{Weisbuch1992}. This favors long-range coherence,  and 
yields condensates that are more robust than without a cavity~\cite{Carusotto:2013gh}.
In this Letter we examine the influence of a resonant planar microcavity on condensation phenomena in 2D semiconductor structures  with unequal electron and hole densities --- see  Fig.~\ref{fig:cartoon}(a-b).
We find  strong matter-light coupling favors small
pairing-momentum states over FFLO states with larger pairing momentum --- specifically it induces breached pair states and anisotropic crescent states, explained below, which spontaneously break both rotational and time-reversal symmetry.
The anisotropic states place excess carriers in a compact crescent-shaped sliver in momentum space 
on the edge of the region occupied by electron-hole pairs, 
as illustrated in Figs.~\ref{fig:cartoon}(c-d), instead of spreading them isotropically.  
The crescent and breached pair states arise only because of coupling to light, and are stabilized by the small photon mass.
Further, as discussed later, the anisotropy also requires long-range Coulomb interactions.   
As such, while the electron-hole-photon model we will introduce below is superficially similar
to the two-channel model of ultracold fermionic atoms~\cite{Giorgini08:RMP}, there are crucial differences:  
For atoms, interactions are contact-like and, most importantly,
the analogue of the photon is a ``closed channel'' molecular state, with a mass twice that of the atoms.  
In addition, phase separation in real space dominates the phase diagram of cold atoms~\cite{zwierlein2006fermionic,partridge2006pairing,shin2006observation}.  The states we propose here are therefore unique to polaritonic systems.  The new crescent states can be identified experimentally by strongly anisotropic 
electrical transport characteristics that can be reoriented by altering the polariton-confinement landscapes 
or by weak resonant optical excitation.  In the following we first explain the calculations that allow us to predict the 
crescent states, and then discuss 
properties that could identify them experimentally.

{\em  Model---}  We consider a model of electrons and holes confined in two-dimensional (2D) quantum wells,
subject to Coulomb interactions, and coupled to  cavity photons.  
The Hamiltonian is thus ($\hbar=1$, $4\pi\varepsilon_0=1$):
\begin{multline}
\label{eq:H}
\hat{H} = \sum_{\k} \left[ \left( \frac{k^2}{2m_e} + E_G \right) \hat{e}_{\k}^{\dagger} \hat{e}^{\mathstrut}_{\k} +
                        \frac{k^2}{2m_h} \hat{h}_{\k}^{\dagger} \hat{h}^{\mathstrut}_{\k}  \right]  
\\ 
      + \frac{1}{2S} \sum_{\k,\k
      ^\prime,\vec p} V_{\vec p}
      \Big{\{}
	  \hat{e}^{\dagger}_{\k+\vec p} \hat{e}^{\dagger}_{\k^\prime-\vec p} \hat{e}^{\mathstrut}_{\k^\prime} \hat{e}^{\mathstrut}_{\k} +
	  \hat{h}^{\dagger}_{\k+\vec p} \hat{h}^{\dagger}_{\k^\prime-\vec p} \hat{h}^{\mathstrut}_{\k^\prime} \hat{h}^{\mathstrut}_{\k} \\ -
	  2 \hat{e}^{\dagger}_{\k+\vec p} \hat{h}^{\dagger}_{\k^\prime-\vec p} \hat{h}^{\mathstrut}_{\k^\prime} \hat{e}^{\mathstrut}_{\k}
      \Big{\}}  + \alpha S (\hat{n}_c - n_0)^2  
\\ 
+ \sum_{\k} \omega_{\k} \hat{a}^{\dagger}_{\k} \hat{a}^{\mathstrut}_{\k}
+ \sum_{\k,\vec p}\frac{g_{\k} }{\sqrt{S}}  \left( \hat{e}_{\k}^{\dagger} \hat{h}_{\vec p-\k}^{\dagger} \hat{a}^{\mathstrut}_{\vec p} + \hat{a}^{\dagger}_{\vec p} \hat{h}^{\mathstrut}_{\vec p-\k} \hat{e}^{\mathstrut}_{\k} \right),
\end{multline}
where $S$ is the system area.  The first term in $\hat{H}$ describes non-interacting electrons and holes with masses $m_{e}$ and $m_{h}$ in a two-dimensional semiconductor 
with band gap $E_G$.
The second term is the mutual Coulomb interaction ${V_\p = 2 \pi e^2 / \varepsilon p}$, while the third term gives the dependence of the electrostatic energy on the system  charge density.
Here $\alpha=e^2 S/2C$ is an (intensive) capacitive scale, which depends on the gating geometry. The target charge density, $n_0$, is proportional to a tunable gate voltage. 
Typically $\alpha$ is large compared to the corresponding interaction scale  ($e^2 n_{e}^{-1/2}/\varepsilon$) so that the actual charge imbalance which minimizes the free energy is nearly identical to the target charge density, i.e. $\langle \hat{n}_c \rangle \simeq n_0$, where
\begin{equation}
\hat{n}_c = \frac 1 S \sum_\k \left( \hat{e}^{\dagger}_\k \hat{e}^{\mathstrut}_\k - \hat{h}^{\dagger}_\k \hat{h}^{\mathstrut}_\k \right) = \hat{n}_e - \hat{n}_h.
\end{equation}
 Including the electrostatic energy realistically, as we do in Eq.~\eqref{eq:H}, allows us to use the 
grand-canonical ensemble without generating unphysical phase separations, and thereby allows us to consider more general variational ansatz states.
The final line of Eq.~\eqref{eq:H} accounts for the cavity photons and their coupling to electrons and holes.
We assume a single branch of cavity photons, and approximate the  dispersion as quadratic, 
$\omega_\k = \omega_0 + {k^2}/{2m_{ph}}$, with typical  mass $m_{ph} \simeq 10^{-4} m_e$.
In the following we measure
lengths in units of the 2D exciton Bohr radius $\aB=\varepsilon/(2\mu e^2)$, where $\mu=m_em_h/(m_e+m_h)$, and energies in units of
$\EB=1/(2\mu\aB^2)$.

To avoid the ultraviolet divergences produced by a momentum-independent matter-light 
coupling~\cite{BCS_polariton, polariton_MacDonald,Kamide_Ogawa_PRL, Kamide_Ogawa_PRB,Levinsen2019}, we take $g_{\k} = g_0 e^{- |\k|/\kappa}$, and choose $1/\kappa$ to be of the order of the material lattice constant.  
This cutoff breaks the theory gauge invariance  under the replacement $\hat{e}_{\vec{k}} \to \hat{e}_{\vec{k} + e\vec{A}}, \hat{h}_{\vec{k}} \to \hat{h}_{\vec{k}-e\vec{A}}$, which could be recovered by taking  $\kappa \to \infty$ and renormalizing the photon 
frequency; see Refs.~\cite{Levinsen2019,Supp}. 
Full gauge invariance requires consistency of the band and 
matter-light coupling Hamiltonians~\cite{Andolina2019}, and is 
crucial to recover the no-go theorems precluding ground state superradiance~\cite{rzazewski75,Andolina2019}.

To control the excitation density we introduce a chemical potential $\mu_{ex}$, and replace
$\hat{H} \to \hat{H} - S \mu_{ex} \hat{n}_{ex}$, 
where 
\begin{equation}
 \hat{n}_{ex} = \frac{1}{S} \sum_{\k} \left[ \hat{a}^{\dagger}_\k \hat{a}^{\mathstrut}_\k + \Frac{1}{2}\left(\hat{e}^{\dagger}_\k \hat{e}^{\mathstrut}_\k + \hat{h}^{\dagger}_\k \hat{h}^{\mathstrut}_\k  \right) \right]. 
\end{equation}
The energy shift accounts for the time-dependence of the non-equilibrium condensates that form at 
finite excitation density.  The no-go theorem does not apply for a system at finite excitation density~\cite{eastham01}.  
We note that because we make the rotating wave approximation, equal shifts in $\omega_0,E_G$ and $\mu_{ex}$ have no effect.

{\em Variational approach---}
To estimate the \emph{finite temperature} phase diagram of our model, 
we use a variational ansatz for the density matrix~\cite{kleinert2009path}, $\hat{\rho}_{v} = \exp(-\beta \hat{H}_v)/\mathcal{Z}_v$,  $\mathcal{Z}_v=\text{Tr}[\exp(-\beta \hat{H}_v)]$. We then minimise the free energy corresponding to this density matrix, $F_v=\langle \hat{H} \rangle_v +   k_B T \, \text{Tr}[\hat{\rho}_v \ln \hat{\rho}_v] = \langle \hat{H} - \hat{H}_v\rangle_v - k_B T \ln \mathcal{Z}_v$, where
$\langle \hat{X} \rangle_v = \text{Tr}(\hat{\rho}_v \hat{X})$.
Standard thermodynamic identities allow one to show that $F_v$ is an upper bound on the true free energy. 
\begin{comment}
based on the inequality $F \equiv - k_B T \ln \Tr \exp(- \beta H)
\leq \mathcal{F}_{\text{vMF}} = F_{\text{MF}} + \langle H - H_{\text{MF}} \rangle_{\text{MF}}$ 
\cite{Stat_mech_Feynman}.  Here $F_{\text{MF}} = - k_B T \ln \Tr \exp \left( - \beta H_{\text{MF}} \right)$  
and $H_{\text{MF}}$ is a general mean-field Hamiltonian whose parameters are fixed by free-energy minimization.  
Minimizing with respect to the variational Hamiltonian $H_{\text{MF}}$ 
can be thought of as minimizing with respect to the parameters of a 
variational ansatz for the density matrix, $\rho_{MF} = \exp(-\beta H_{\text{MF}})$.
Minimising with respect to $\mathcal{F}_{\text{vMF}}$ then corresponds to finding the optimal density-matrix which 
is diagonal in a ( to be determined ) occupation number representation.  
\end{comment}
%
The variational Hamiltonian $\hat{H}_v$ should be chosen to be solvable, and for our model, we should allow for 
electron-hole coherence, photon coherence, population imbalance, and arbitrary polariton momentum $\Q$.
We therefore consider a variational Hamiltonian of the form:
\begin{multline} 
\label{MF_var_hamilt}
\hat{H}_{v} = 
  \nu_{\Q} \sqrt{S} \phi ( \hat{a}^{\dagger}_\Q  +  \hat{a}^{\mathstrut}_\Q )  +  
  \sum_\q
      \nu_\q \hat{a}^{\dagger}_\q \hat{a}^{\mathstrut}_\q  \\  + \sum_\k 
      \begin{pmatrix}
	  \hat{e}_{\frac{\Q}{2} + \k}^{\dagger} & \hat{h}_{\frac{\Q}{2} - \k}^{\mathstrut}
      \end{pmatrix}
      \begin{pmatrix}
	  \eta_{\k}^e &  \Delta_{\k}  \\ 
	  \Delta_{\k}  & -  \eta_{\k}^h
      \end{pmatrix}
      \begin{pmatrix}
	  \hat{e}_{\frac{\Q}{2} + \k}^{\mathstrut} \\ 
	  \hat{h}_{\frac{\Q}{2} - \k}^{\dagger}
      \end{pmatrix}.
\end{multline}
We can derive an expression for $F_v$ in terms of the eigenvalues and 
eigenstates of  $\hat{H}_{v}$ (see supplemental material~\cite{Supp}). The first term in Eq.~\eqref{MF_var_hamilt} is chosen so that the photon density is $\phi^2$. The results  below are then obtained by minimizing  
over  the variational parameters ($\phi, \nu_\q, \eta^e_\k, \eta^h_\k, \Delta_\k, \Q$).  
Because this ansatz contains only pairing of fermions and displacement of bosons, it is equivalent to mean field theory approaches.

{\em Pairing phases---}  Previous work~\cite{Varley_Lee_PRB2016} explored the ground-state phase 
diagram of Eq.\eqref{eq:H} in the absence of coupling to photons, using the grand canonical ensemble with a charge imbalance 
chemical potential $\mu_c$ ($\hat{H} \to \hat{H} - \mu_c S \hat{n}_c $) in place of a realistic electrostatic 
energy
\footnote{Ref.~\cite{Varley_Lee_PRB2016} however neglects intraspecies interactions which may affect its conclusions, see~\cite{Subashi2010}}.
It predicted first order phase transitions between a balanced condensate
with $\langle \hat{n}_c \rangle = 0$ and  an imbalanced $\langle \hat{n}_c \rangle \neq 0$ anisotropic FFLO condensate with 
non-zero center-of-mass momentum $Q \sim |\langle \hat{n}_{e}\rangle^{1/2} - \langle \hat{n}_{h}\rangle^{1/2}|$.
When applied to the exciton only problem, our more realistic description of electrostatics 
shows that the transition between a $\Q=\0$ condensate and the FFLO state (see Ref.~\cite{Supp}) is continuous as a 
function of gate voltage.  

When the balanced condensate is coupled to photons, it becomes a polaritonic state, with exciton--photon
coherence, further lowering its energy.
In contrast, coupling to photons has little influence on the FF  state because 
excitons with center of mass momentum $\Q$ couple to photons at the same momentum, and the small photon mass
places these far off resonance.  The photon fraction in the FF state is therefore very small, and we thus refer to this state as dark.
Coupling to photons  therefore favors states with a small center of mass momentum. 
Numerical minimization indeed reveals that, at moderate imbalance, coupling to photons 
yields a  bright polaritonic condensate state with $\Q$  small but non-zero. 
Surprisingly, this state accommodates  excess charged carriers by spontaneously breaking rotational and 
time reversal symmetry.   At larger imbalance, the expected FF phase is recovered --- for the extremely charge imbalanced case, see Ref.~\cite{tiene_2019}.

\begin{figure}[htpb]  
\includegraphics[width=1.0\linewidth]{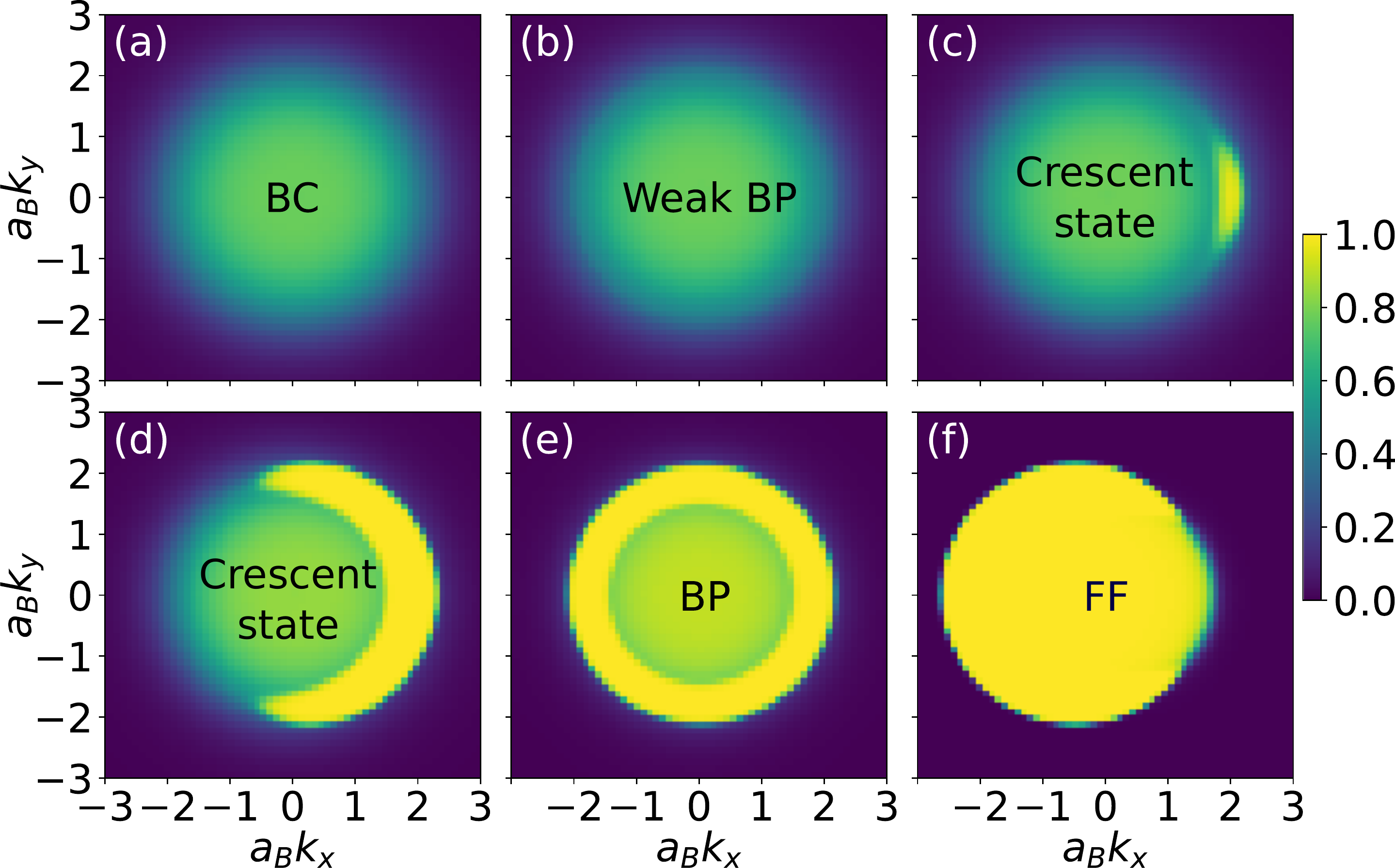}
\caption{
Electron occupation $\langle \hat{e}^{\dagger}_{\vec{Q}
/2 + \vec{k}} \hat{e}^{}_{\vec{Q}/2 + \vec{k}} \rangle$ 
\units{
for various imbalance values $n_0 \aB^2$: (a) 0, (b) $6.25\times 10^{-3}$, (c) $1.875 \times 10^{-2}$, (d) $0.125$, (e) $0.1875$, (f) $0.25$. 
} Labels on each panel indicate the phases as described in the text. %
\units{
The values of $Q \aB$ are (c) $0.5 \times 10^{-6}$, (d) $0.5 \times 10^{-5}$, (f) $1.05$,
} and zero for panels (a),(b),(e). Other parameters are as in  Fig.~\ref{fig:cartoon}.
}
\label{fig:state_vs_n0}
\end{figure}

Figure~\ref{fig:state_vs_n0} shows how the electron momentum distribution changes with charge imbalance --- corresponding cross sections showing also hole occupation and coherence are presented in~\cite{Supp}.
Panel (a) shows the case with $n_0=0$, i.e. balanced populations. At small $n_0$ (panel (b)), the state maintains $\Q=\0$ to take optimal advantage of the photon-mediated electron-hole coupling.  In the zero temperature limit,
%, unless phase separation occurs,
accommodating extra charges requires forming a Fermi surface, which encloses regions of momentum space in which both valence and 
conduction band states are occupied.  At low charge imbalance, the Fermi sea forms a ring at the outer edge of the region of paired electrons. 
We will refer to the state at low carrier densities
as a ``weak breached pair'' (WBP) state, as it is reminiscent of the two-Fermi surface breached pair state described in 
Ref.~\cite{Forbes2005:BP}. In contrast to the fully breached pair, the coherence in 
Fig.~\ref{fig:state_vs_n0}(b), is only weakly suppressed in the region where extra electrons exist because the 
temperature is comparable to the conduction band Fermi energy.  
For intermediate values of $n_0$, illustrated in panels (c,d), we find a surprising broken rotational symmetry anistropic 
state with $0 < Q \ll |\langle \hat{n}_{e}\rangle^{1/2} - \langle \hat{n}_{h}\rangle^{1/2}|$.  
The unpaired carriers in this state are contained in a Fermi pocket with a crescent shape on the edge of the otherwise circular electron distribution, hence we refer to it as the crescent state (CS).  
As $n_0$ increases further, the crescent extends in angle.  Eventually it is replaced by a filled annulus (panel e),  equivalent 
to the breached pair (BP) state of Ref.~\cite{Forbes2005:BP}, and related to the Sarma state~\cite{sarma1963influence}. 
Finally, at large enough $n_0$, one recovers the dark FF state.   Further increasing $n_0$ brings the system to a normal state (not shown).  This sequence occurs at high excitation density. At low excitation density (not shown) the BP state is replaced by a Sarma state where excess particles occupy a single isotropic Fermi surface~\cite{sarma1963influence}, matching the extreme imbalance limit~\cite{tiene_2019}.

\begin{figure}[ht]  
\includegraphics[width=1.0\linewidth]{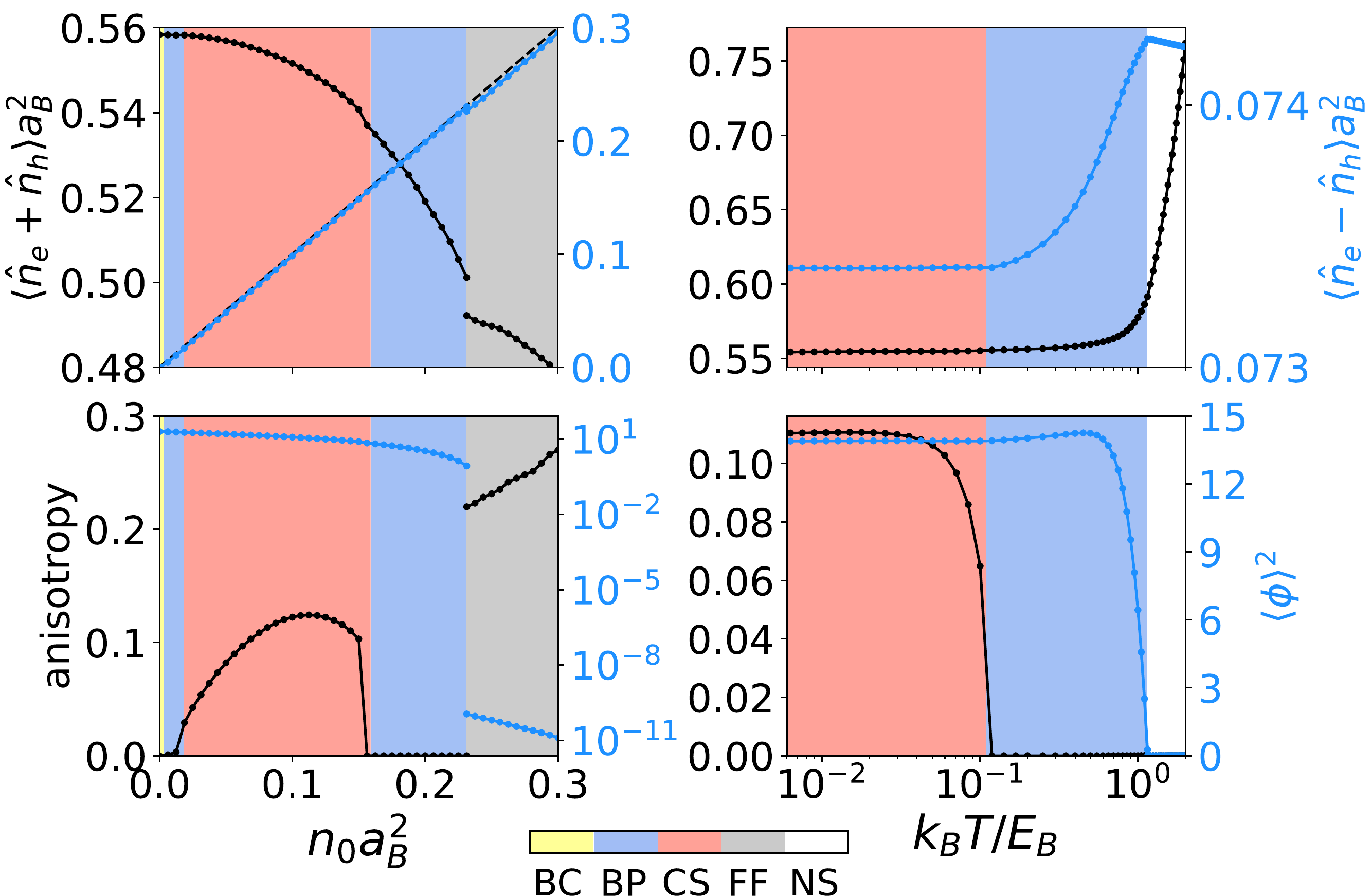}
\caption{%
\units{Evolution of state with target charge density $n_0$ at $k_B T = 0.04 \EB$ (left) and with temperature $T$ at $n_0 \aB^{2} = 0.075 $ (right);}
other parameters as in Fig.~\ref{fig:cartoon}.  
Top panels show excitonic density (black; left axis) and charge imbalance (blue; right axis). The dashed blue line shows $n_0$. Bottom panels show anisotropy (black; left axis) and photon density $\phi^2$  (blue; right axis).}
\label{fig:F_phi_dn_anis_vs_n0}
\end{figure}

{\em Phase diagram ---} 
Figure~\ref{fig:F_phi_dn_anis_vs_n0} illustrates how the 
minimum free energy state evolves with target charge density and temperature by plotting 
charge imbalance, electronic excitation density, photon density, and anisotropy
$\mathcal{A} \equiv \sum_{\vec{k}}| \hat{\vec{k}} \cdot \hat{\vec{Q}} |\langle \hat{e}^{\dagger}_{\vec{Q}/2 + \vec{k}} \hat{e}^{}_{\vec{Q}/2 + \vec{k}} \rangle / \sum_{\vec{k}} \langle \hat{e}^{\dagger}_{\vec{Q}/2 + \vec{k}} \hat{e}^{}_{\vec{Q}/2 + \vec{k}} \rangle$.  
This figure demonstrates that that the crescent state
persists over a wide temperature range, before being replaced by the weakly breached pair (isotropic) state.  
From this figure we see that most transitions, other than those into and out of the $BP$ state are continuous.

\begin{figure}[ht]  
\includegraphics[width=1.0\linewidth]{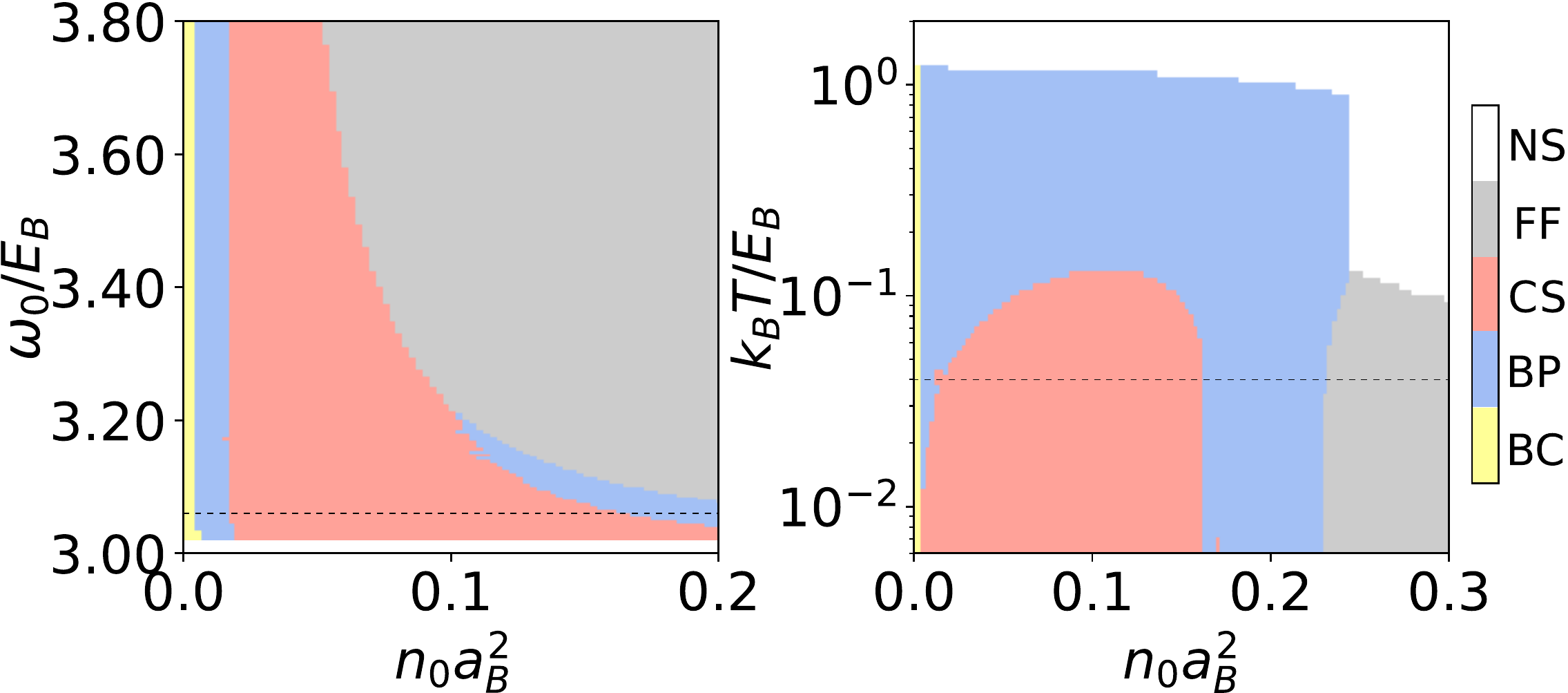}
\caption{
Phase diagrams.  
\units{Left: vs charge density $n_0$ and photon cutoff frequency $\omega_0$ at $k_B T=0.04 \EB$. }
\units{Right:  vs charge density $n_0$ and temperature $T$ at $\omega_0 = 3.06 \EB$. The dashed lines indicate $\omega_0 = 3.06\EB$ (left) and $k_B T = 0.04 \EB$ respectively.}
All other parameters are as in Fig.~\ref{fig:cartoon}.}
\label{fig:ph_diagr}
\end{figure}

The quantities plotted in  Fig.~\ref{fig:F_phi_dn_anis_vs_n0} allow us to 
classify phases, and extract the phase diagrams in Fig.~\ref{fig:ph_diagr}. 
Because the BP and crescent states have significant photon fractions,
the small photon mass should allow them to survive to high temperature even when the collective fluctuations 
(absent in our mean-field theory) are included~\cite{Keeling2005}.  
In contrast, the excitonic FF state should be restricted to low temperatures, 
%i.e. those typical for exciton condensation, 
due to the larger excitonic mass.  
Since the crescent state is stabilised by the matter-light coupling,
an experimentally accessible way to alter its robustness
is by changing the photon cutoff frequency, $\omega_0$, {\it e.g.}, using a wedge cavity.
When the photon is detuned far above the exciton energy, the cavity plays little role and excitonic results should be recovered.  Figure~\ref{fig:ph_diagr}(a) shows such a phase diagram, {\it vs.} $n_0$ and $\omega_0$.  
Because physical states require $\mu_{ex} < \omega_0$, the lower boundary of this phase diagram 
cuts off just above this limit.
As expected the crescent state becomes less prominent with increasing $\omega_0$, although a narrow
stability interval persists up to large detunings.

{\em Crescent State Properties---} 
The crescent state is anisotropic, like the FF state, but has a significant photon fraction and a qualitatively smaller 
pairing momentum.  Because of its anisotropy, it is not immediately clear whether it has zero net current as 
expected by Bloch's theorem~\cite{Bohm49:Bloch}.  An explicit calculation shows that the crescent state 
has a non-zero excitonic current (electron current plus hole current) that is balanced
by an  equal and opposite photon current --- i.e. a counterflow condensate state --- generated by a shift in the condensate pair momentum 
from $\Q=\0$ to $\Q_{min} \ne 0 $~\footnote{We note that in principle a similar statement, that a non-zero photon current and exciton current exist, but cancel at the optimum $\Q$, also holds for the FF state.
However as the FF state is almost entirely dark, this photonic current is negligible.}.  
The momentum shift
balances matter energy gain against photon kinetic energy cost.
Since the shift is small enough to leave the electron and hole distributions almost unchanged, we can approximate
$\Q_{min} \simeq(m_{ph}/|\phi|^2 )[ \sum_{\k} \vec k [ \langle \hat{e}^{\dagger}_{\k} \hat{e}^{\mathstrut}_{\k}\rangle/m_e + \langle \hat{h}^{\dagger}_{\k} \hat{h}^{\mathstrut}_{\k} \rangle/m_h]$, {\it i.e.},
$|\Q_{min}|$ is parametrically small due to the small photon to electron mass ratio.
Indeed, as noted in the caption of Fig.~\ref{fig:state_vs_n0}, our numerical results for 
$|\Q_{min}|$ in the crescent state are orders of magnitude smaller than in the FF state.

Since Bloch's theorem~\cite{Bohm49:Bloch} can be generalized to a coupled photon-matter system,
we expect that the charge current (electron current minus hole current) also vanishes.  
In our numerical calculations, we find that this cancellation is imperfect, but ascribe the non-zero
numerical result to the UV matter-light coupling  cutoff $\kappa$ discussed previously.
In the supplemental material~\cite{Supp}, we show that this charge current 
vanishes as the UV cutoff diverges.  

The crescent state is a metal with a Fermi surface for unpaired electrons, and we expect that it 
will exhibit metallic transport properties.  
The anisotropic Fermi surface in Fig.~\ref{fig:state_vs_n0} implies 
anisotropic electrical transport with larger conduction along the thin direction of the 
crescent, {\it i.e.} in the direction parallel to $\Q$, that can be used to identify the crescent state 
experimentally.  
Any weak perturbation, for example weak resonant excitation or spatial anisotropy of a weak 
polariton confinement landscape, can be used to control the sense of anisotropy - possibly {\it in situ.}
Also, since the crescent state breaks inversion symmetry, nonlinear {\it ac} response is also 
expected to exhibit rectification.

Notably, both strong matter-light coupling and long-range Coulomb interactions are required to stabilize the CS crescent state. While the photon promotes $\Q \approx \0$ pairing, it is the long-range Coulomb interaction which favors anisotropy.
Indeed, screening the Coulomb interaction eventually leads to a continuous transition from the 
anisotropic crescent state to an isotropic state (see  Ref.~\cite{Supp}).  
We therefore expect that our mean-field calculations overestimate the stability range of the crescent state.

{\em Conclusions---}
Since the crescent and breached pair unbalanced states are polaritonic, 
they are expected to survive to high temperatures and should therefore be accessible in current experiments involving doped quantum wells~\cite{Brunhes_PRB1999,Rapaport_PRB2001,Qarry_SST2003,Bajoni_PRB2006,Gabbay_PRL2007,Smolka_Science2014} or two-dimensional materials in cavities~\cite{Sidler_NP2016,chakraborty2018control,fernandez2019electrically}. 
Our work focuses on the small imbalance regime where we are most confident about our conclusions.
At high doping, one instead may consider Fermi-edge (Mahan) excitons, see e.g.~\cite{mahan2013:many,Pimenov_PRB2017} and refs.\ therein.  Open questions include how the states we consider here connect to these Fermi-edge states, the 
effects of electronic screening in a charge doped system, and practical treatments that go beyond mean-field theory.

{\em Acknowledgments---}
We acknowledge helpful discussions with J.~Levinsen, M.~Parish, and P.~Pieri.
The Flatiron Institute is a division of the Simons Foundation. 
AS acknowledges support from the EPSRC CM-CDT (EP/L015110/1) and a travel award from the Scottish Universities Physics Alliance.  AS, AHM and JK acknowledge financial support from a Royal Society International Exchange Award, IES\textbackslash{}R2\textbackslash{}170213.
FMM acknowledges financial support from the Ministerio de Econom\'ia y Competitividad (MINECO), project No.~MAT2017-83772-R. 
JK acknowledges financial support from EPSRC program ``Hybrid Polaritonics'' (EP/M025330/1).  AHM acknowledges support from Army Research Office (ARO) Grant \# W911NF-17-1-0312 (MURI).  
This work was performed in part at Aspen Center for Physics, which is supported by 
National Science Foundation grant PHY-1607611 and 
partially supported by a grant from the Simons Foundation.

\bibliographystyle{apsrev4-1_abbrv}
\bibliography{literature}

\onecolumngrid
\clearpage

\renewcommand{\theequation}{S\arabic{equation}}
\renewcommand{\thefigure}{S\arabic{figure}}
\setcounter{equation}{0}
\setcounter{figure}{0}
\setcounter{page}{1}

\section{Supplementary Material for: ``Crescent states in charge-imbalanced polariton condensates''}
\twocolumngrid

\section{Variational free energy}

We provide here explicit expressions for the variational free energy, $F_v$.  
To derive this, we note that diagonalising the Hamiltonian in Eq.~\eqref{MF_var_hamilt} requires a shift of the photon operator, $\hat{a}_\Q \to \hat{a}_{\Q} - \sqrt{S} \phi$.  The fermionic part of the Hamiltonian is diagonalised by a unitary transform:
\begin{equation}
\begin{pmatrix}
\hat{e}_{{\Q}/{2} + \k}^{} \\ 
\hat{h}^{\dagger}_{{\Q}/{2} - \k}
\end{pmatrix}
= 
\begin{pmatrix}
u_{\k} &  \upsilon_{\k} \\ 
-\upsilon_{\k} & u_{\k}
\end{pmatrix}
\begin{pmatrix}
\hat{c}_{+,\k}^{} \\ 
\hat{c}_{-,\k}^{\dagger}
\end{pmatrix} ,
\end{equation} 
where 
\begin{subequations}
\label{u_and_v_functions}
\begin{align}
u_{\k} &= \sqrt{\frac 1 2 \left( 1 + \frac{\eta_{\k}^e + \eta_{\k}^h }{2 E_{\k}} \right)}, \\ 
\upsilon_{\k} &= - \sign(\Delta_{\k})
		    \sqrt{\frac 1 2 \left( 1 - \frac{\eta_{\k}^e + \eta_{\k}^h  }{2 E_{\k}} \right)}, \\ 
E_{\k} &= \sqrt{\left( \frac{\eta_{\k}^e + \eta_{\k}^h }{2} \right)^2 + \Delta_{\k}^2 }.
\label{eq:spectrum}
\end{align}
\end{subequations}
We define the resulting eigenvalues of the diagonalised Fermionic problem as
\begin{equation}
    \varepsilon_{\k}^{\pm}=E_{\k} \pm (\eta^e_{\k} - \eta^h_{\k})/2.
\end{equation}
Using this,  expressions such as the electron 
$N_{\k}^e = \langle \hat{e}_{\k + \vec Q / 2}^{\dagger} \hat{e}_{\k + \vec Q / 2}^{} \rangle$ and hole 
$N_{\k}^h =\langle \hat{h}_{\vec Q / 2 - \k}^{\dagger} \hat{h}_{\vec Q / 2 - \k}^{} \rangle$ populations  can be expressed in terms of variational parameters via: 
\begin{align*}
    N^e_\k &= u^2_\k n_F(\varepsilon^+_\k) + \upsilon^2_\k [ 1 - n_F(\varepsilon^-_\k) ],\\
    N^h_\k &= \upsilon^2_\k [ 1 - n_F(\varepsilon^+_\k)] + u^2_\k n_F(\varepsilon^-_\k),
\end{align*}
where $n_F(\varepsilon)$ is the Fermi distribution.
Because the variational state is Gaussian, the expectations of quartic terms in the Hamiltonian can be decoupled via Wick's theorem.  
When putting this all together, we will first take the continuum (large $S$ limit), where momentum sums become integrals.  Then, as described in the main text, we use $\aB=\varepsilon/(2\mu e^2)$ as a lengthscale, and so introduce a dimensionless momentum $\tilde{\k}=\aB \k$. We thus find:
\begin{widetext}
\begin{multline}
\label{variat_free_energy}
\Frac{F_v}{S \aB^{-2}} = 
- \int\!\frac{d\tilde{\k}}{(2 \pi)^2} 
  \Bigg{[}
    \varepsilon^{+}_{\k} n_F(\varepsilon^{+}_{\k}) + \frac{1}{\beta} \ln \left( 1 + e^{-\beta \varepsilon^{+}_{\k}} \right) + (\varepsilon^{+}_{\k} \to \varepsilon^{-}_{\k})
  \Bigg{]} 
  +
    \frac{\alpha}{\aB^2}
    \left\{ 
                                        \iint \frac{d\tilde{\k} d\tilde{\k}'}{(2\pi)^4} N^c_\k  N^c_{\k^\prime} -  2 n_0 \aB^2  \int\! \frac{d\tilde{\k}}{(2\pi)^2} N^c_{\k}
                                    \right\}
    \\ + \frac{E^X_{ee} + E^X_{hh}}{S \aB^{-2}}
       - \iint \frac{d\tilde{\k} d\tilde{\k}'}{(2\pi)^4}
                    \Bigg{\{} 
                    \frac{V_{\k-\k^\prime}}{\aB^2}
                    u_{\k} \upsilon_{\k} 
                    \Big{[}
                      1 - n_F(\varepsilon^{+}_{\k}) - n_F(\varepsilon^{+}_{\k})
                    \Big{]} 
                    u_{\k^\prime} \upsilon_{\k^\prime}
                    \Big{[}
                      1 - n_F(\varepsilon^{+}_{\k^\prime}) - n_F(\varepsilon^{-}_{\k^\prime})
                    \Big{]}
                    \Bigg{\}} \\  + 
                            \int \frac{d\tilde{\k}}{(2 \pi)^2}
                            \Bigg{\{}  
                                E^e_{\k + \frac{\Q}{2}} N^e_\k + 
                            	E^h_{\frac{\Q}{2} - \k} N^h_{\k} 
                            \Bigg{\}} 
                                      +\phi^2 \aB^2  (\omega_\Q - \mu_{ex})
  + 
            2\phi \aB
            \int \frac{d\tilde{\k}}{(2 \pi)^2} 
            \frac{g_{\k+\frac{\Q}{2}}}{\aB} u_{\k} \upsilon_{\k} 
            \Big{[}
                n_F(\varepsilon^{+}_{\k}) + n_F(\varepsilon^{-}_{\k}) - 1
            \Big{]}.
\end{multline}
\end{widetext}
Here $N^c_{\k} = N^e_\k - N^h_\k$, the bare electronic energies are:
\begin{subequations}
\begin{align}
\label{el_tot_energy}
E^e_{\k} &= \EB \frac{m_h \tilde{k}^2}{m_e + m_h} 
+ E_G - \frac 1 2 \mu_{ex}, \\
E^h_{\k} &= \EB \frac{m_e \tilde{k} ^2}{m_e + m_h} 
 - \frac 1 2 \mu_{ex},
\label{hole_tot_energy}     
\end{align}
\end{subequations}
written in terms of $\EB=1/(2 \mu \aB^2)$
and the exchange energies 
\begin{equation}
\frac{E^X_{ee/hh}}{S \aB^{-2}} = 
    - \frac 1 2 \iint \frac{d\tilde{\k} d\tilde{\k}^\prime}{(2 \pi)^4}  N^{e/h}_\k \frac{V_{\k-\k^\prime}}{\aB^2} N^{e/h}_{\k^\prime}.
\end{equation}

We may note that in Eq.~\eqref{variat_free_energy}, the quantities
$n_0 \aB^2$ and $\phi \aB$ are dimensionless, while $\alpha/\aB^2$, $V_\p /\aB^2$ and $g_{\p}/\aB$ have units of energy as expected.  For the Coulomb interactions, we may define:
\begin{equation}
    v_{\tilde{\p}} \equiv \frac{V_{\p}}{\aB^2} = \frac{2 \pi \EB}{\tilde{p}}
\end{equation}
Since we include the global electrostatic energy explicitly, we use a definition where $V_{\0}$ is set to zero.

\subsection{Numerical evaluation}

At each point in the numerical minimisation, one must evaluate the energy and its derivatives.
Calculation of the expectation of the Coulomb interactions requires a 4D integral of the form 
\begin{displaymath}
I_C = \iint d \tilde{\k} d \tilde{\k}^\prime f(\tilde{\k})  v_{\tilde{\k}-\tilde{\k}^\prime} f (\tilde{\k^\prime}).
\end{displaymath}
Evaluating this on a grid of $N\times N$ points would require $N^4$ operations, significantly limiting the values of $N$ that can be used. However,  by rewriting this integral one can significantly reduce the computational effort involved.  
Using the Fourier transform $\tilde{f}(\tilde{\x}) \equiv  \int {d\tilde{\k}} f(\tilde{\k}) e^{i\tilde{\k}\cdot \tilde{\x}}/{(2\pi)^2}$, 
one can rewrite $I_C$ as a 2D integral in real space 
$I_C = \int d\tilde{\x} |\tilde{f}(\tilde{\x})|^2 \tilde{v}(\tilde{\x})$, which now requires only $N^2$ operations.  
At the same time, Fast Fourier Transform of a 2D function requires $O(N^2 \ln N)$ operations. 
Therefore, calculating the integral $I_C$ in real space reduces the scaling of the number of operations 
from $N^4$ to $N^2 \ln N$ allowing to do the full 2D optimisation efficiently 
on a reasonable momentum grid, e.g. for $10^4$ $k$-points. 
To implement the above idea, in the second line of Eq.~\eqref{variat_free_energy} we 
define $f(\tilde{\k}) = u_{\k} \upsilon_{\k} 
[1 - n_F(\varepsilon^{+}_{\k}) - n_F(\varepsilon^{+}_{\k})]$.
Exchange energies can be rewritten in the same way:
$E^X_{ee/hh} = - \frac 1 2 \int d\x |\tilde{N}^{e/h}(\x)|^2 \tilde{v}(\textbf{x})$.

We numerically implement the minimisation by using the truncated Newton algorithm from the SciPy~\cite{SciPy} library.   As local minima can exist in the free energy landscape (see below), we use a method equivalent to an adiabatic sweep.  Specifically, as we vary a control parameter, we use the optimal variational parameters found for one value of the control parameter as initial conditions for the minimisation at the next value of the control parameter.  As discussed further below, where there can be hysteresis, we use repeated sweeps with increasing and decreasing control parameters.

\section{Electron, hole and coherence cross sections at  \texorpdfstring{$k_y = 0$}{ky=0}}
\label{sec:state_vs_n0__ky_0}

\begin{figure}[ht]  
\includegraphics[width=1.0\linewidth]{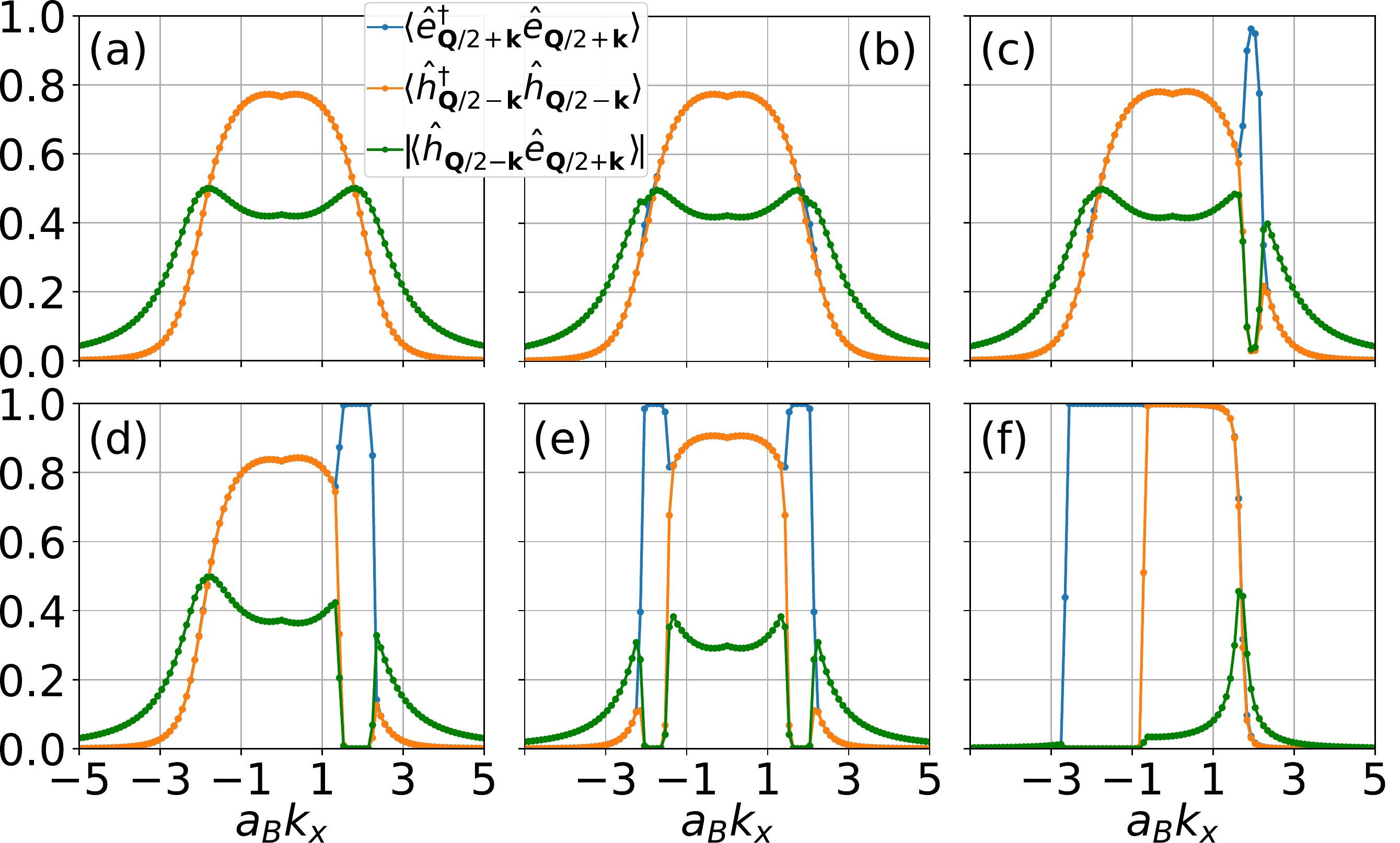}
\caption{Cross sections at $k_y=0$, showing the electron and hole populations, and electron-hole coherence.  Panels shown here correspond to those shown in Fig.~\ref{fig:state_vs_n0} of the main text. }
\label{fig:state_vs_n0__ky_0_cut}
\end{figure}

\begin{figure}[ht]  
\includegraphics[width=1.0\linewidth]{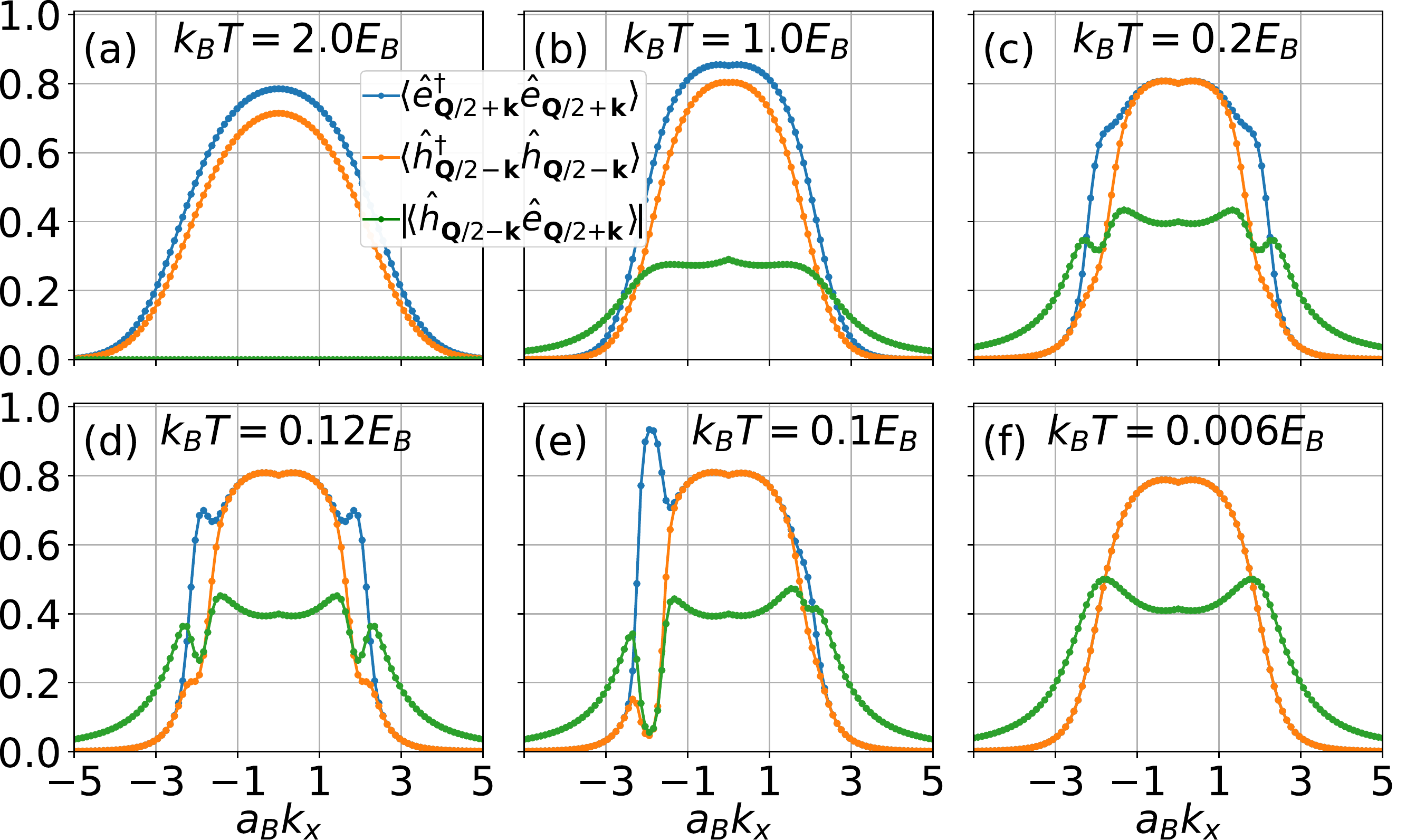}
\caption{Cross sections at $k_y=0$ for various temperatures as indicated, 
%for charge density $n_0 \aBO^2=0.3$. 
\units{for charge density $n_0 \aB^2=0.075$. }
All other parameters as in Fig.~\ref{fig:cartoon} of the main text. }
\label{fig:state_vs_T__ky_0_cut}
\end{figure}

Figures~\ref{fig:state_vs_n0__ky_0_cut} and~\ref{fig:state_vs_T__ky_0_cut} show cross sections of the electron and hole densities and the coherence function at varying charge density and temperature. In plotting these figures, the rotational symmetry breaking is always chosen such that there is symmetry  about the $k_y=0$ line, so that any crescent will intersect the cross section shown. As also discussed in the caption of Fig.~\ref{fig:cartoon} of the main text, these cross sections show that within the crescent or ring, both conduction and valence bands are occupied, while at other momenta, a total of one band is filled, so that electron and hole populations are equal.     At low enough temperatures, the crescent or breached pair states show a complete suppression of the hole population within the Fermi surface.  When the temperature becomes comparable to the conduction band Fermi energy, the suppression is weaker.

\begin{comment}
\begin{figure}[ht]  
\includegraphics[width=0.95\linewidth]{ph_diagr_om0_n0__50x50_k_points__om0_153}
\caption{Phase diagram vs charge density $n_0$ and temperature $T$ at $\omega_0 = 1.53$. All other parameters as in Fig.~\ref{fig:cartoon}.}
\label{fig:n0_T_ph_diagr}
\end{figure}
\end{comment}

\section{First-order phase transitions}
\label{sec:CC_BP_1st_ord_ph_tr}
\begin{figure}[ht]  
\includegraphics[width=1.0\linewidth]{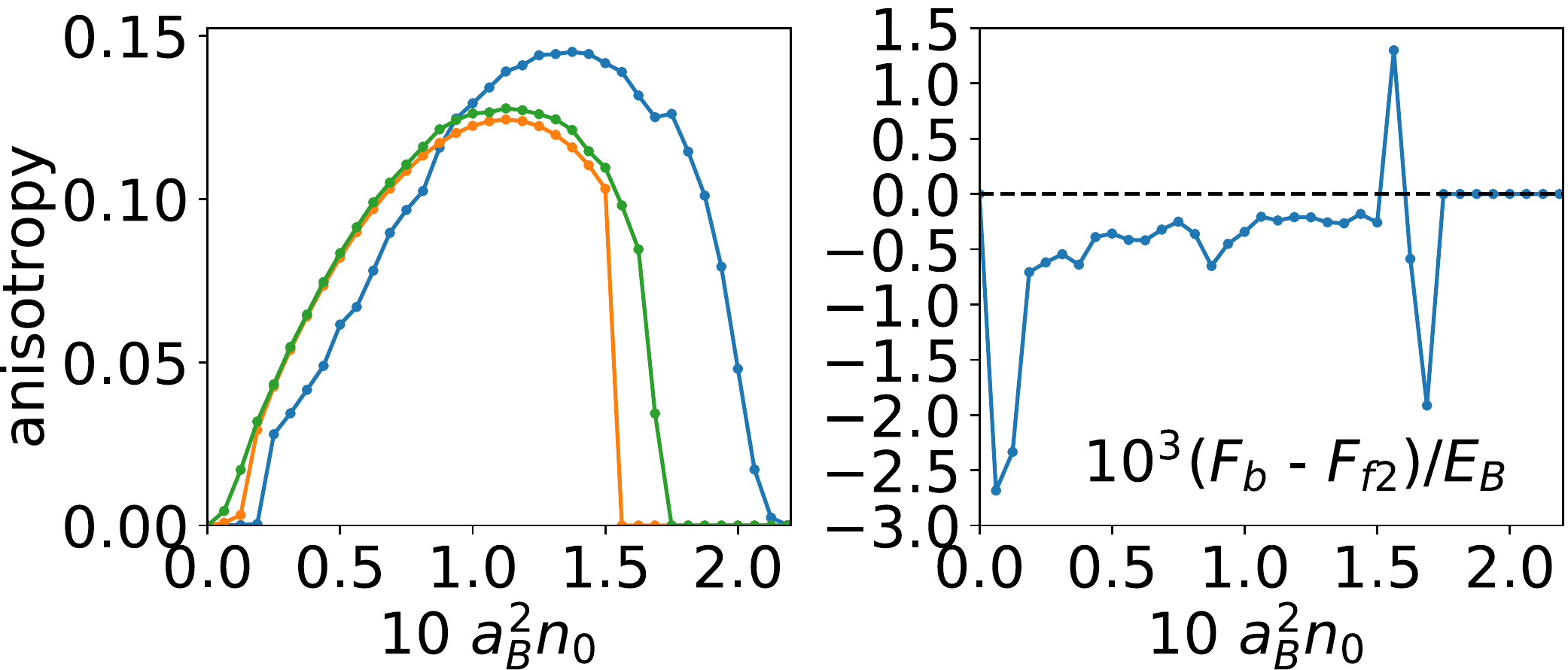}
\caption{Hysteresis associated with the counterflow condensate to breached pair transition.  Data is collected using adiabatic sweeps of charge density (i.e. using minimum found at the previous value of $n_0$ as the initial guess for the next value of $n_0$).  Three sweeps are shown; first increasing $n_0$ (f1), then decreasing $n_0$ (b) and then again increasing $n_0$ (f2).  The first forward sweep did not achieve a global minimum so should be discarded;  the subsequent sweeps do find consistent solutions for most $n_0$, but hysteresis is seen around the CS-BP transition. All other parameters are as in Fig.~\ref{fig:cartoon}.}
\label{fig:increasing_decreasing_n_0}
\end{figure}

As seen from the evolution of the anisotropy order parameter with charge imbalance shown in Fig.~\ref{fig:F_phi_dn_anis_vs_n0}, the transition from the crescent state (CS) state to the breached pair (BP) state is discontinuous.  This indicates the transition is first order, associated with the existence of two distinct local minima of the free energy.  Figure~\ref{fig:increasing_decreasing_n_0} shows that corresponding to this, one sees hysteresis in the anisotropy, as measured by comparing an adiabatic sweep of increasing vs decreasing the charge density $n_0$.

\begin{figure}[ht]  
\includegraphics[width=1.0\linewidth]{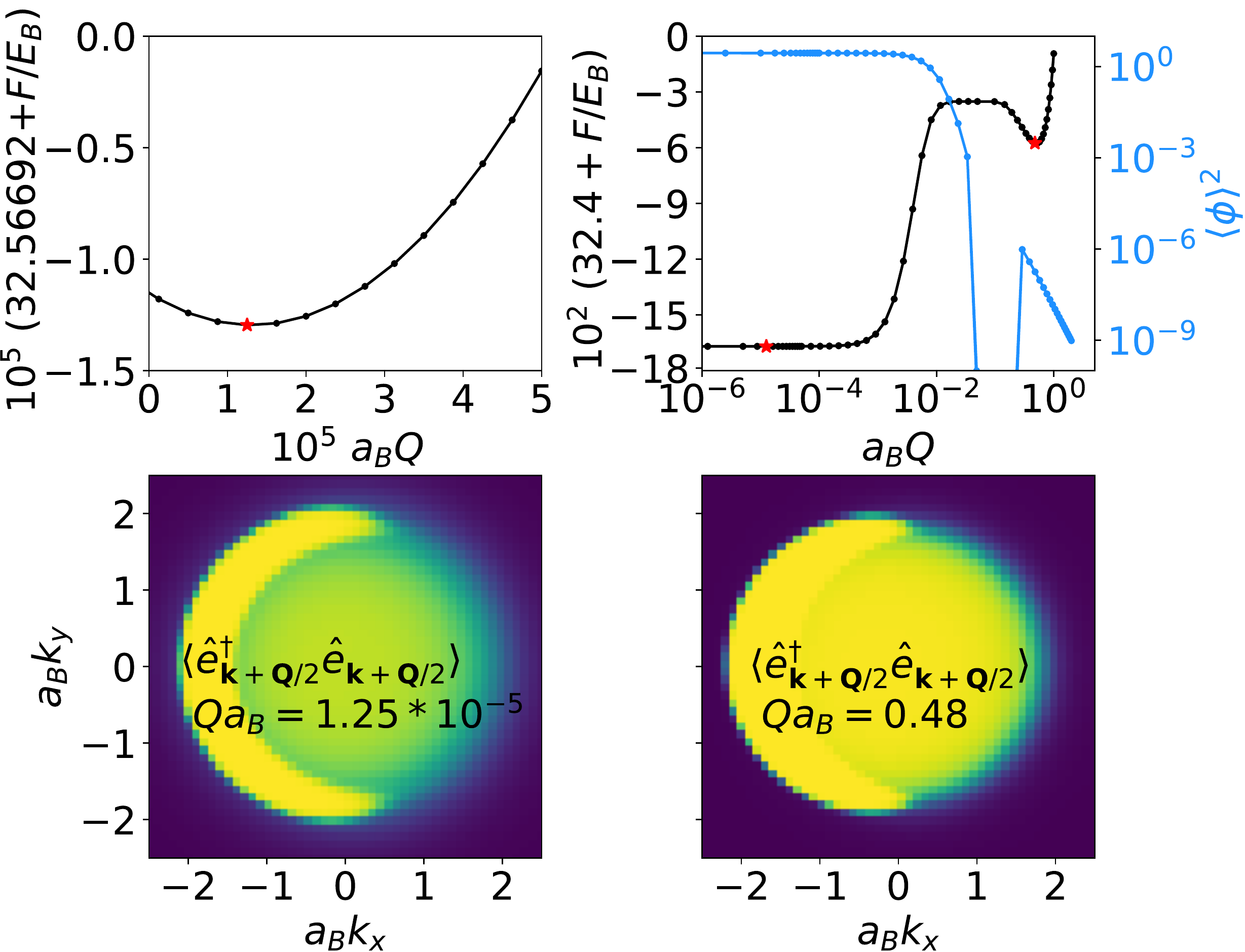}
\caption{Coexisting local minima corresponding to counterflow condensate and FFLO states.  Top panels show the free energy and photon order parameter vs pairing wavevector $Q$.  The top left panel is an expanded region of small $Q$ around the optimal wavevector for the CS state.  The bottom two panels show the electron mode occupation for the CS (left) and FF (right) solutions, corresponding to the red stars shown in the top panels.  
%Parameters are $n_0 \aBO^2 = 0.4$, $\omega_0 = 1.55 E_G$, 
\units{Parameters are $n_0 \aB^2 = 0.1$, $\omega_0 = 3.1 \EB$, }
and all other parameters as in Fig.~\ref{fig:cartoon}}
\label{fig:fflo_vs_crescent}
\end{figure}

As well as the existence of separate local minima for the CS and BP state, one can also find parameter regimes where the CS and FF state solutions exist as competing local minima.  Indeed, as seen from the phase diagram, Fig.~\ref{fig:ph_diagr}, at large photon energy $\omega_0$, there is a direct CS-FF transition.  Figure~\ref{fig:fflo_vs_crescent} illustrates this, showing the free energy landscape vs $Q$ and the solutions corresponding to the two local minima.

\section{Gauge invariance}
\label{sec:cutoff_and_gauge_inv}
As mentioned in the Letter, the exponential momentum cutoff $\kappa$ for the matter-light interaction regularise %removes 
the UV divergence, but breaks gauge invariance. In this section we discuss the consequence of this and a possible way to restore the gauge invariance by considering $\kappa \to \infty$ and renormalising the photon frequency~\cite{Levinsen2019}.  We focus here on the invariance under transformations involving a static and uniform change of the vector potential, as these are sufficient to understand the issues introduced by the cutoff $\kappa$.  A more complete discussion of the necessity of gauge invariant models when considering ground-state phase transitions can be found in Ref.~\cite{Andolina2019}.

Consider a simple gauge transformation by adding a constant vector-potential to the electron and hole momenta, $\k \to \k \pm e \A$, where $e$ is the electronic charge, and the sign of shift depends on the type of quasiparticle.  After this shift the kinetic energy part of the Hamiltonian then becomes:
\begin{equation}
    \sum_{\k}
    \frac{1}{2m}
    \Bigg{\{} 
        \left(\k + e\A \right)^2 \hat{e}^{\dagger}_{\k} \hat{e}^{\mathstrut}_{\k} + \left(\k - e\A \right)^2 \hat{h}^{\dagger}_{\k} \hat{h}^{\mathstrut}_{\k}
    \Bigg{\}}.
\end{equation}
To be gauge invariant, the model must be invariant under this transformation. 
Relabeling operators $\hat{e}_{\k} \to \hat{e}_{\k+e\A}$ and $\hat{h}_{\k} \to \hat{h}_{\k-e\A}$ clearly recovers the original kinetic part of the Hamiltonian. 
One can also readily check that this relabeling does not affect the Coulomb term.  However, it does change the matter-light interaction term 
$\sum_{\k} g_{\k}^{\mathstrut} \hat{e}^{\dagger}_{\k} \hat{h}^{\dagger}_{\q-\k} \hat{a}_{\q}^{\mathstrut}$, as this now becomes:
\begin{equation}
    \sum_{\k} g_{\k}^{\mathstrut} \hat{e}^{\dagger}_{\k+e\A} \hat{h}^{\dagger}_{\q-\k-e\A}\hat{a}_{\q}^{\mathstrut} = 
    \sum_{\k} g_{\k-e\A}^{\mathstrut} \hat{e}^{\dagger}_{\k} \hat{h}^{\dagger}_{\q-\k}\hat{a}_{\q}^{\mathstrut}.
\end{equation}
One clearly sees that the momentum dependence of the coupling constant makes the model gauge dependent.

The total charge current can be related to the derivative of the free energy with respect to vector potential, i.e. ${\bf j_A}=d F/d \A$.
Since the free energy $F$ of a gauge-invariant model cannot depend on a constant gauge shift, the charge current in such a case is identically zero.
However, since our model breaks the gauge invariance, $F(\A)$ has a minimum at a non-zero value of $\A$, which implies a finite charge current at $\A=\0$, ${\bf j_0}=dF/d\A|_{\A=\0}$.

\begin{figure}[ht]  
\includegraphics[width=1.0\linewidth]{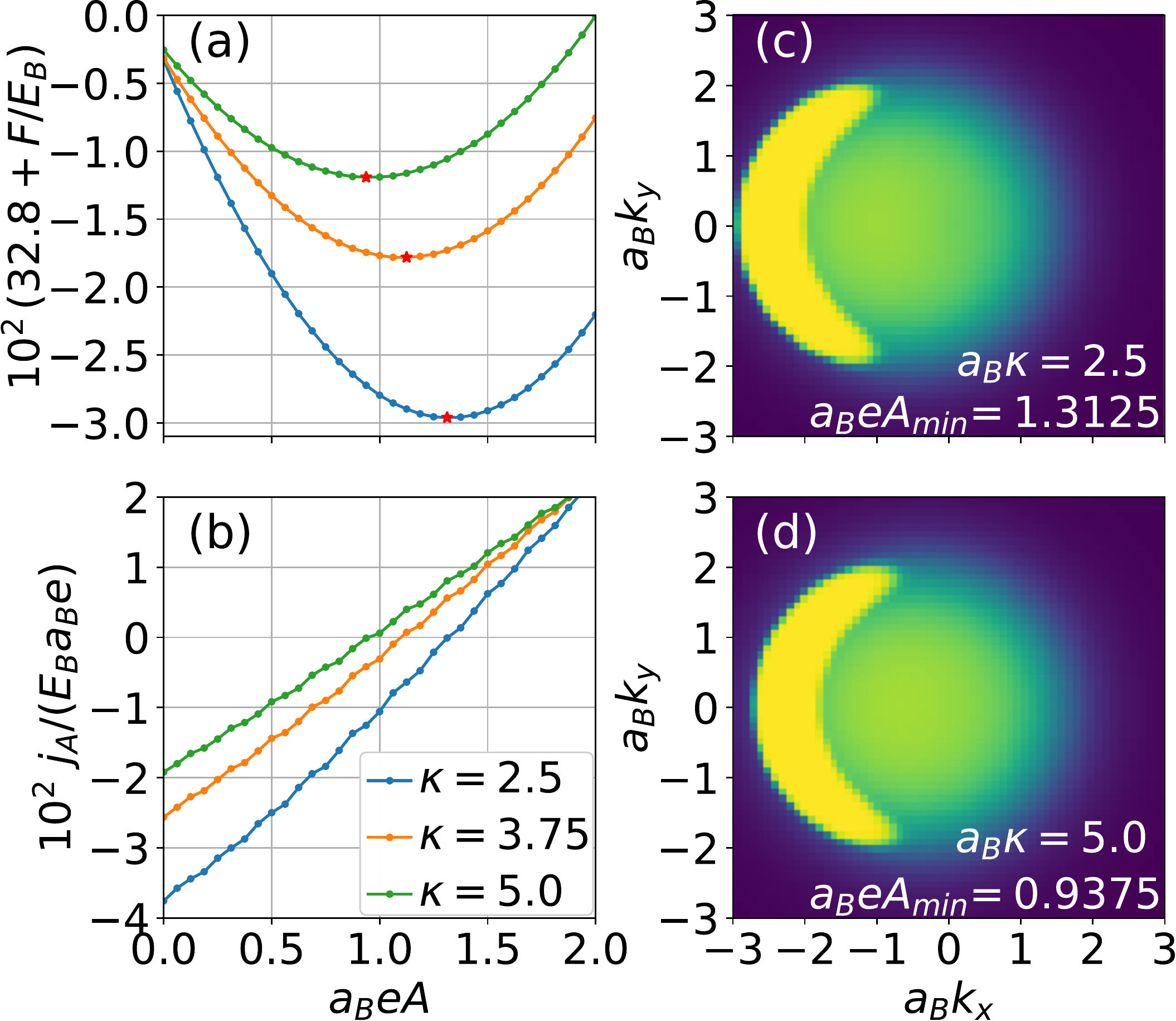}
\caption{(a) Free energy $F(\A)$ versus constant vector potential shift $\A$ and (b) corresponding charge current,  ${\bf j_A}=dF(\A)/d\A$.  
\units{The three lines correspond to photon cutoff frequency and momentum cutoffs $\omega_0 = 3.06 \EB, \kappa = 2.5 \aB^{-1}$ (blue), $\omega_0 = 3.1055 \EB, \kappa = 3.75 \aB^{-1}$ (orange) and $\omega_0 = 3.1435 \EB, \kappa = 5.0\aB^{-1}$ (green).}
(c,d) electron occupations at the positions of the free energy minima 
\units{for $\kappa=2.5 \aB^{-1}$ and $\kappa=5.0 \aB^{-1}$. Target charge density $n_0 \aB^2 = 0.1$;}
all other parameters are the same as in the main part of the paper.}
\label{fig:Gauge_question}
\end{figure}

To recover gauge invariance within our model, one needs  a momentum independent coupling $g_\k$, or equivalently, to send the cut-off to 
infinity, $\kappa \to \infty$.  This introduces an ultraviolet divergence, however, as shown in Ref.~\cite{Levinsen2019}, this divergence can be removed  renormalising the bare photon frequency.  Following Ref.~\cite{Levinsen2019}, one can show that  to keep the renormalised photon frequency constant under a change in the momentum cutoff from $\kappa_1$ to $\kappa_2$ requires a shift of the bare photon energy $\omega_0$. At large $\kappa$ this shift is approximately given by $\delta \omega_0 \approx (g_0^2 \mu / \pi) \ln(\kappa_2 / \kappa_1)$.
% JK:  In the above, I think that it's simplest to write the actual
% shift, in dimensionful units.
In Fig.~\ref{fig:Gauge_question} we plot $F(\A)$ for various values of the cutoff, 
%$\kappa \aBO = 5.0, 7.5, 10.0$.
\units{$\kappa \aB = 2.5, 3.75, 5.0$.}
In plotting this figure, we in fact choose the value $\omega_0$ at each cutoff $\kappa$ so as to ensure that the free energy at $A=0$ remains constant.  Note that this means that in this figure we implicitly use a renormalisation scheme where the shift of $\omega_0$ is chosen for the non-zero value of $n_0$ used in this figure.
The bare frequencies used are given in the caption.

From Fig.~\ref{fig:Gauge_question} we see firstly that the overall scale of the current (or equivalently the variation of the $F(\A)$ with $\A$) reduces with increasing cutoff.  Moreover, in the right panels of this figure we plot the electron mode occupations at the minimum of $F(\A)$, $A_{min}$, for 
%$\kappa\aBO = 5.0, 10.0$.
\units{$\kappa\aB = 2.5, 5.0$.}
It is clear that changing the cutoff does not significantly change the electron distribution. These results   suggest that, in the limit $\kappa \to \infty$, by renormalising the bare photon 
frequency, we obtain gauge invariant results, which remain qualitatively the same as those we found with a finite cutoff.

\section{Second-order BC-FF phase transition without a photon, \texorpdfstring{$m_e = m_h$}{me=mh}}
\begin{figure}[ht]  
\includegraphics[width=1.0\linewidth]{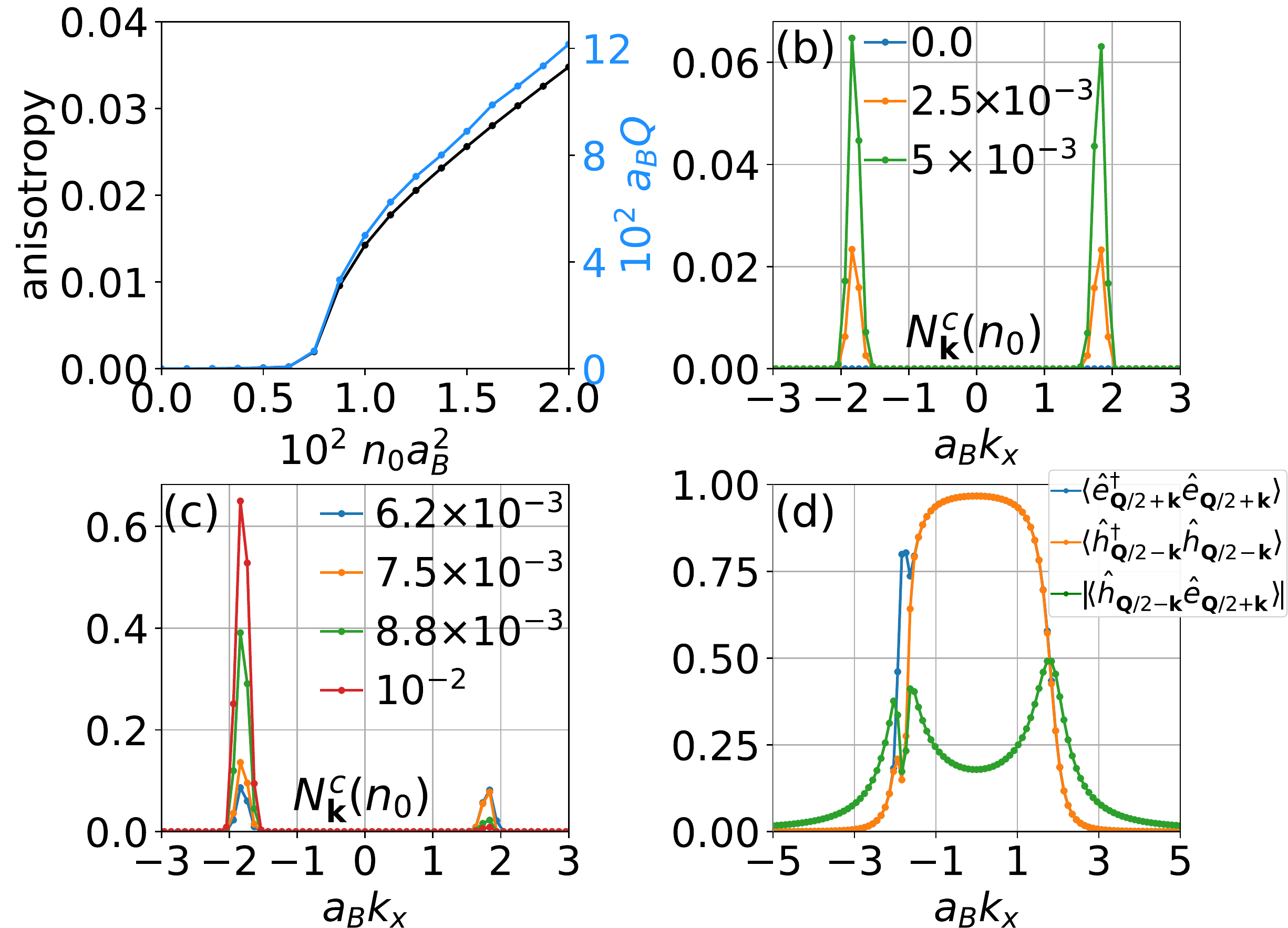}
\caption{Second order phase transitions for the excitonic limit (no coupling to light). 
(a) Evolution of anisotropy (black, left) and center of mass momentum $Q$ (blue, right)  vs increasing density imbalance $n_0$. (b) Momentum resolved net charge distribution $N_{\k}^c=\langle \hat{e}^{\dagger}_{\Q/2 + \k} \hat{e}^{}_{\Q/2 + \k} - \hat{h}^{\dagger}_{\Q/2 - \k} \hat{h}^{}_{\Q/2 - \k} \rangle$  at small imbalance, showing a weakly BP state.
(c) Same quantity at larger imbalance, showing the appearance of the FF state. 
(d) Electron and hole occupation, coherence at 
%$n_0 \aBO^2 = 0.04$ 
\units{$n_0 \aB^2 = 0.01$}
--- just into the FF state. Note the opposite momentum offset for electron and hole states.}
\label{fig:fflo_2nd_opt}
\end{figure}

In this section we present the behaviour of the purely excitonic system, using the explicit gating scheme we consider to fix the charge density.  Previous work~\cite{Varley_Lee_PRB2016}  found a first order transition between a balanced condensate phase and an FFLO state, by working in the grand canonical ensemble and thus introducing a chemical potential for imbalance, $H \to H - \mu_{c} S n_c $.  However, Ref.~\cite{Varley_Lee_PRB2016} also neglected intraspecies interactions, i.e. electron-electron and hole-hole repulsion.  \citet{Subashi2010} in contrast found that including such intraspecies interactions makes phase transitions continuous.
Here, we show that with our explicit gating process and including intraspecies interactions, even in the absence of photons, we  indeed observe a second order transition, with  the pairing $Q$-vector growing continuously as density imbalance increases --- see Fig.~\ref{fig:fflo_2nd_opt}. As shown in the figure, we in fact find that, within our gating model,
a narrow region of a weakly breached pair state exists
between the balanced condensate and FFLO states --- i.e. a state with excess charge density on a ring near the Fermi surface.

\section{Energetic origin of the CS state}

\subsection{Effects of screening on the CS state}

\begin{figure}[ht]  
\includegraphics[width=1.0\linewidth]{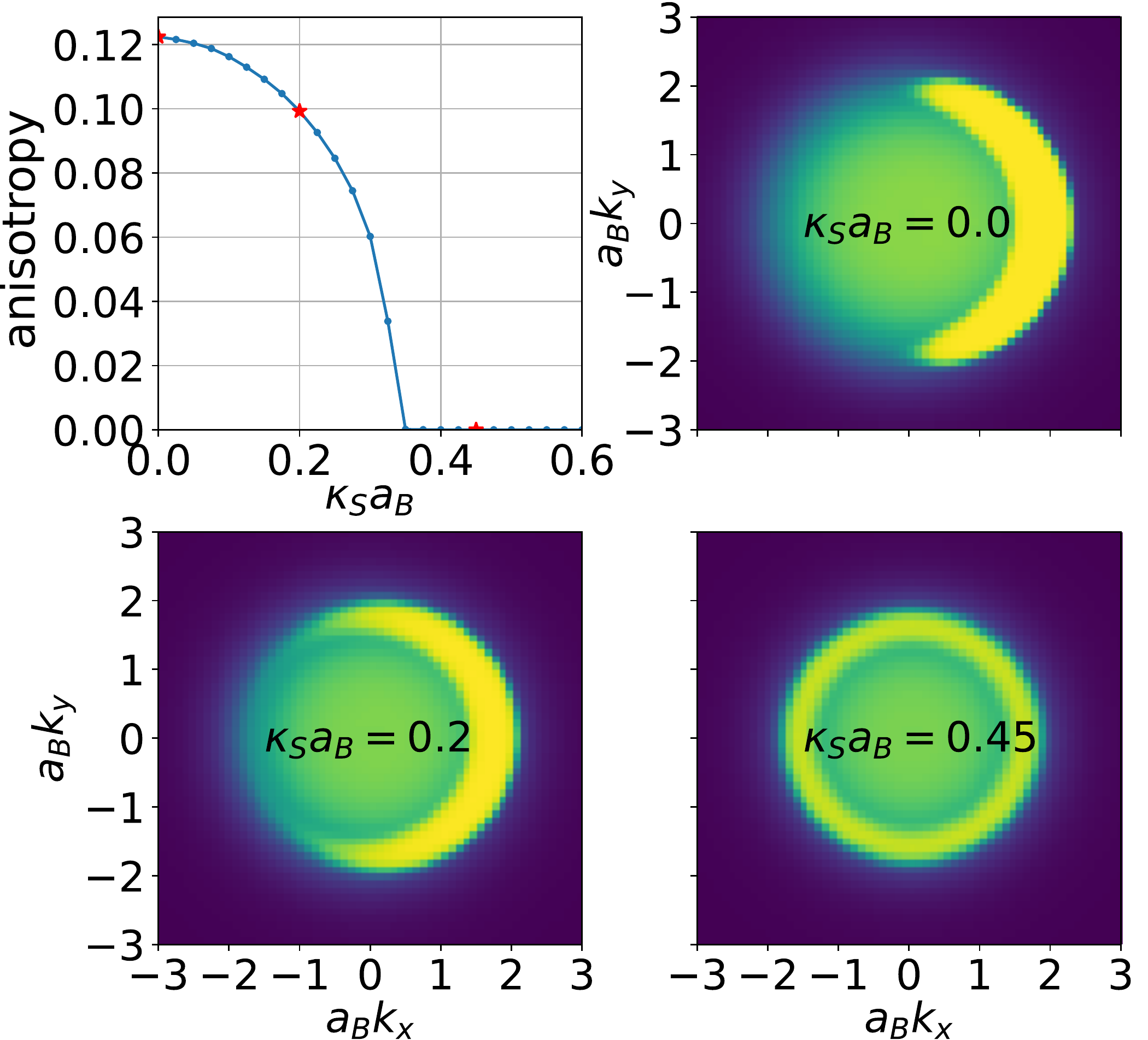}
\caption{ Effect of screening $V_\k(\kappa_S) = 2\pi e^2 / \varepsilon(k + \kappa_S)$; 
%$n_0\aBO^2 = 0.4$, $\omega_0 = 1.53 E_G$, 
\units{$n_0\aB^2 = 0.1$, $\omega_0 = 3.06 \EB$,}
other parameters are the same as in Fig.~\ref{fig:cartoon}.
Top left panel shows the dependence of anisotropy on screening. 
Other panels show electron occupations corresponding to points highlighted by red crosses. }
\label{screening_effect}
\end{figure}

While it is the coupling to light which stabilizes $\Q \simeq \0$ imbalanced states vs FF states, in this section we prove that the anisotropic crescent state also requires long-ranged Coulomb interactions.  To demonstrate this,  Fig.~\ref{screening_effect}  shows how the anisotropy changes as we introduce screening of the
Coulomb interaction. We consider a Yukawa potential, $V_\k(\kappa_S) = 2\pi e^2 / \varepsilon(k + \kappa_S)$, where $1/\kappa_S$ is a screening length, such that $\kappa_S=0$ recovers the unscreened Coulomb interaction.  As seen in Fig.~\ref{screening_effect}, the anisotropy vanishes when the screening length approaches the
bare exciton Bohr radius.  From the colormaps, we see that as anisotropy vanishes, the crescent state is replaced by the breached pair state.

\subsection{Competition of kinetic and Coulomb energies}

\begin{figure}[ht]  
\includegraphics[width=1.0\linewidth]{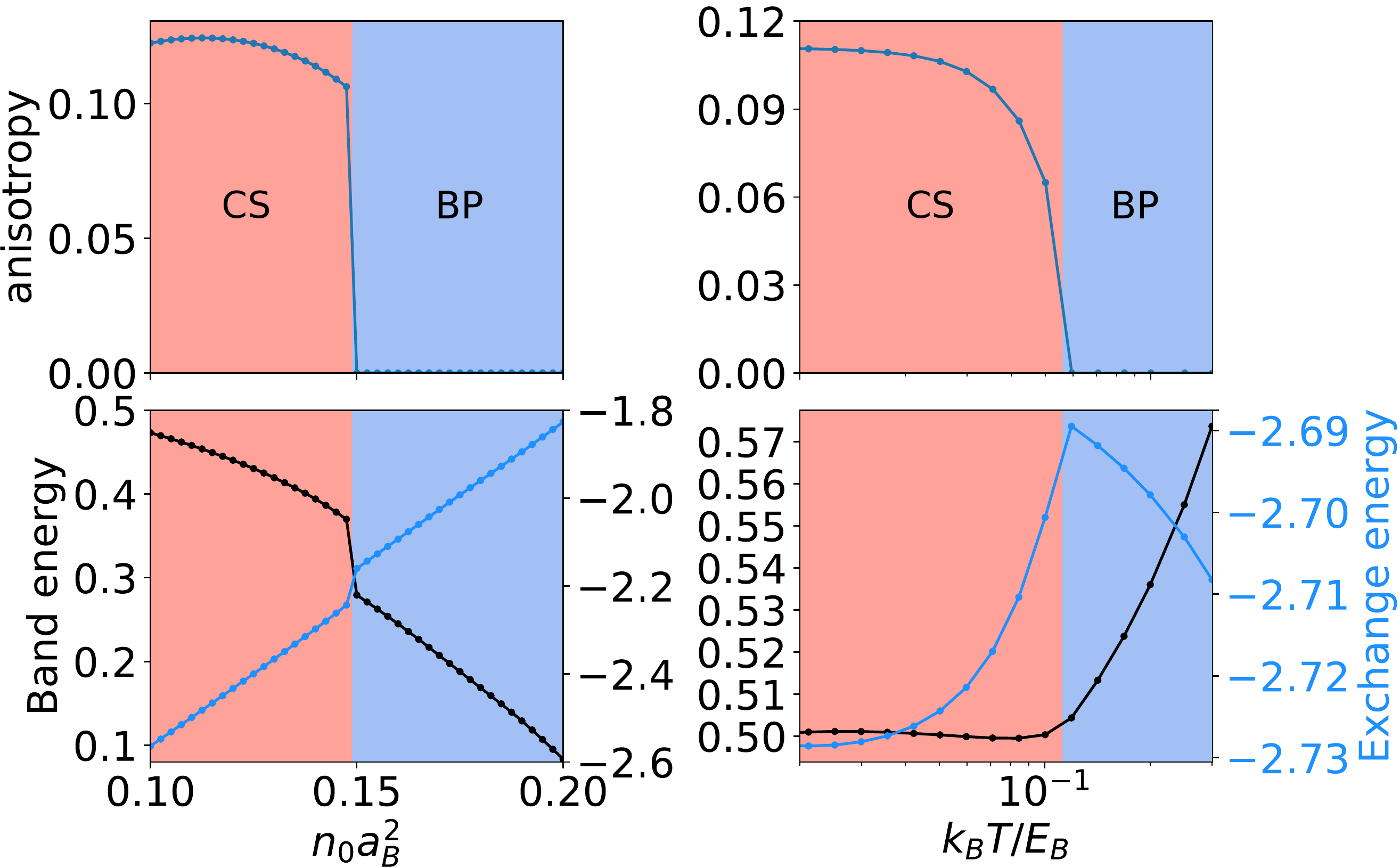}
\caption{Energy decomposition of free energy at the boundary of CS and BP phases.
Top left and right panels show, for reference, anisotropy versus charge density $n_0 \aB^2$ and temperature $T$ respectively.
Bottom panels show corresponding kinetic energy (sum of electron and hole kinetic energies) and exchange energy
(total  of electron-electron, hole-hole and electron-hole terms). 
All parameters are the same as in left and right panels of Fig.~\ref{fig:F_phi_dn_anis_vs_n0} respectively.}
\label{fig:nature_CC_BP}
\end{figure}

Figure~\ref{fig:nature_CC_BP} provides further evidence that the Coulomb interactions are important in driving the formation of the anisotropic phase.  In this figure we show have the electronic kinetic and Coulomb energies vary as we cross the boundary between the CS and BP phase.  For the first order transition with increasing target charge density, we see clearly that the first order boundary is a competition between the CS state, with higher kinetic energy and lower exchange, and the BP state with lower kinetic energy but higher exchange.  This confirms the conclusion above that it is the long-range Coulomb interaction which favours the CS state.

\section{Mass imbalance}
In the Letter, we presented results only for the case where the conduction and valence band have equal masses.
In this section, we show how mass imbalance --- which is usually present in real materials --- affects the CS state.

\begin{figure}[ht]  
\includegraphics[width=1.0\linewidth]{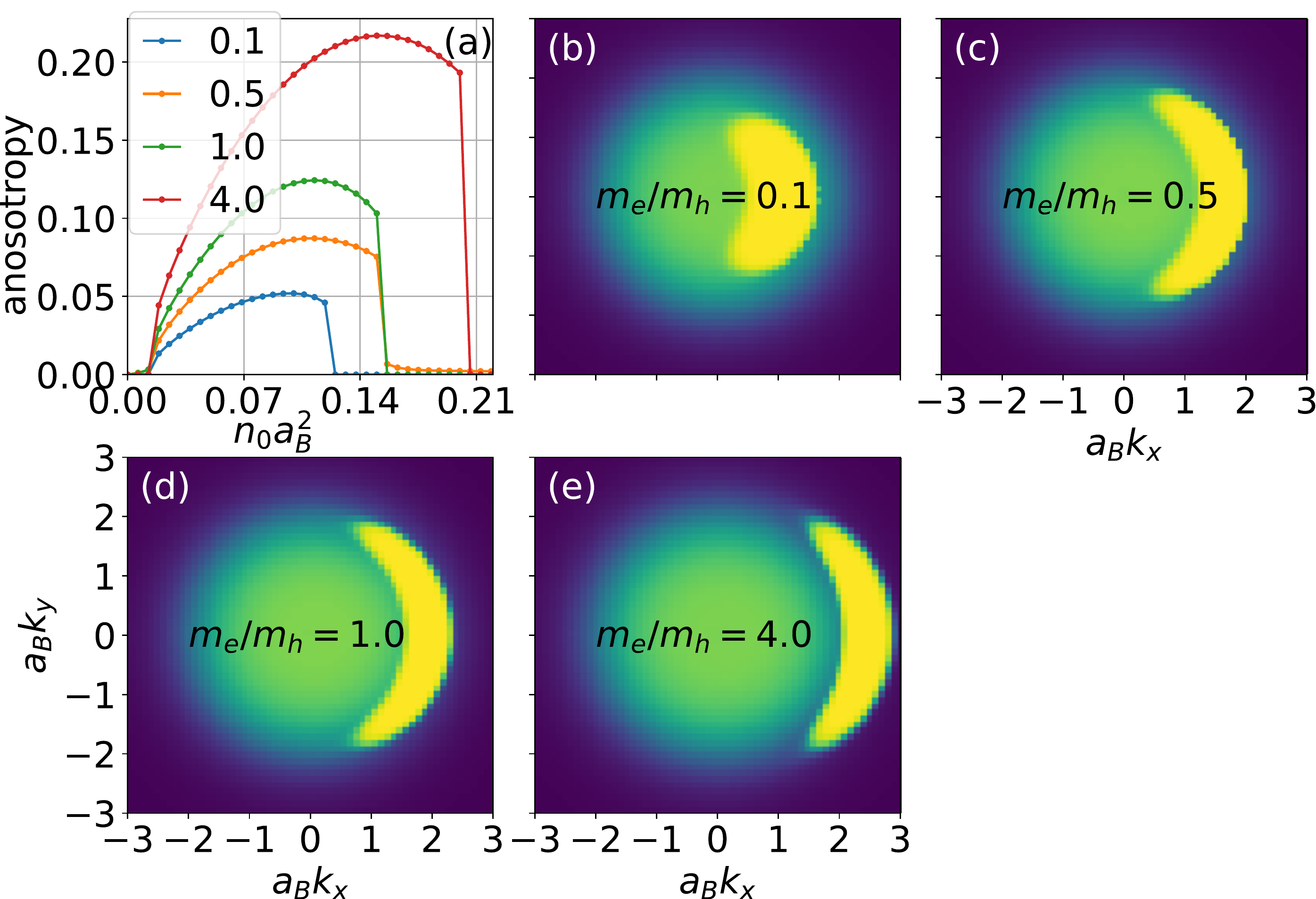}
\caption{(a) Anisotropy versus target charge density $n_0$ at different mass imbalance. 
(b-e) electron occupations at 
%$n_0 \aBO^2 = 0.3$ 
\units{$n_0 \aB^2 = 0.075$ }
corresponding to different mass ratios as labelled.  All other parameters are the same as in Fig.~\ref{fig:cartoon}.}
\label{Q0state_anis_vs_memh}
\end{figure}

In Fig.~\ref{Q0state_anis_vs_memh} we plot the 
dependence of anisotropy on $n_0$, and the electron occupations at %$n_0 = 0.3 \aBO^{-2}$,
\units{$n_0 = 0.075 \aB^{-2}$,} for four values of $m_e/m_h$ ranging from $0.1$ to $4.0$. Typically electron mass is lower than the hole mass, $m_e/m_h<1$.  However,  since our Hamiltonian is
invariant under a transformation $e \leftrightarrow h, n_0 \to -n_0$, one can consider the behaviour for $m_e/m_h > 1$ as indicating the behaviour when there is hole doping rather than electron doping.

Clearly, all results are qualitatively the same, although as seen from Fig.~\ref{Q0state_anis_vs_memh}(a), a reduced mass ratio shrinks the range of $n_0$ where the CS state occurs.  In addition, changing mass ratio distorts the region of momentum space where the extra electrons are found.  Heavier electrons --- Fig.~\ref{Q0state_anis_vs_memh}(e) --- lead to a more extended crescent, while lighter electrons to a less extended one --- Fig.~\ref{Q0state_anis_vs_memh}(b). 

At yet higher target charge densities $n_0$, the system adopts either the FF state or a normal state.  When the mass ratio $m_e/m_h$ becomes small (for electron doping), the FF state becomes less stable, and is replaced by the normal state~\cite{Varley_Lee_PRB2016}.
This comes from the increased separation of electron and hole 
Fermi surfaces when $m_e \ll m_h$ and $n_e > n_h$. In contrast, the opposite mass ratio brings Fermi energies closer.

\end{document}